# U.S. Port Disruptions under Tropical Cyclones: Resilience Analysis by Harnessing Multiple-Source Dataset

Chenchen Kuai
Zachry Department of Civil & Environmental Engineering
Texas A&M University
College Station, TX 77843, USA
Email: mobility@tamu.edu

Zihao Li*, Ph. D.
Zachry Department of Civil & Environmental Engineering
Texas A&M University
College Station, TX 77843, USA
Email: scottlzh@tamu.edu

Yunlong Zhang*, Ph. D.
Zachry Department of Civil & Environmental Engineering
Texas A&M University
College Station, TX 77843, USA
Email: yzhang@civil.tamu.edu

Xiubin Bruce Wang, Ph. D.
Zachry Department of Civil & Environmental Engineering
Texas A&M University
College Station, TX 77843, USA
Email: bwang@civil.tamu.edu

Dominique Lord, Ph. D.
Zachry Department of Civil & Environmental Engineering
Texas A&M University
College Station, TX 77843, USA
Email: d-lord@tamu.edu

Yang Zhou, Ph. D.
Zachry Department of Civil & Environmental Engineering
Texas A&M University
College Station, TX 77843, USA
Email: yangzhou295@tamu.edu

U.S. Port Disruptions under Tropical Cyclones: Resilience Analysis by Harnessing Multiple-Source Dataset

**Abstract**
This study introduces the CyPort Dataset, recording disruptions to 145 U.S. principal ports and freight network from 90 tropical cyclones (2015-2023). It addresses limitations of event specific resilience studies and provides a comprehensive dataset for broader analysis. To account for excess zeros and unobserved heterogeneity in disruption outcomes, the Random Parameter Negative Binomial Lindley (RPNB Lindley) model is employed to produce more reliable resilience insights. The model demonstrates improved fit over traditional methods and uncovers variation in how features such as wind speed, storm surge height, rainfall, and distance to cyclone influence disruption outcomes across ports. This analysis reveals a tipping point at Saffir Simpson Hurricane Category 4, where disruptions escalate sharply, causing greater impacts and prolonged recovery. Regionally, ports along the Gulf of America show greatest vulnerability. Within the freight network, ports with high betweenness centrality are more resilient, while transshipment and local hubs are more fragile.

# 1. Introduction

Ports serve as critical hubs in freight transportation and are especially vulnerable to extreme weather events due to their coastal locations. The disruptions from extreme weather events can lead to operational shutdowns of the ports, widespread delays across the freight shipping network, and in severe cases, infrastructure damage. Among these events, tropical cyclones[1] (TCs) are particularly disruptive. During the 2020 Atlantic hurricane season, ten named cyclones made landfalls in the United States, causing extensive port closures and an estimated 41.1 billion dollars in supply chain disruptions (Chambers et al., 2024). The critical role of ports in freight transportation, combined with the disruptions caused by TC, highlights the importance of port resilience.

Port resilience refers to the capacity of the port systems to withstand and adapt to adverse conditions, and recover positively from the disruptions (Qin et al., 2023; United Nations, 2022). Supported by the availability of Automated Identification System (AIS) data (i.e., vessel trajectories) and comprehensive extreme weather records, data-driven approaches have become increasingly popular to better understand port resilience under the impact of TCs (Bläser et al., 2024; Farhadi et al., 2016; Wang et al., 2024). These studies often conduct empirical analysis and focus on specific regions affected by a significant TC event to analyze the extent of disruption and the patterns of recovery. Zhang et al., (2024) studied the Tropical Cyclone Veronica's impact on Northwestern Australia ports, detailing vessel level impact caused by the TC, including delayed arrival and extra travel distance between ports. Zohoori et al., (2023) assessed the operational performance of the Houston Ship Channel using AIS data before, during, and after Hurricane Harvey in August 2017. They applied the number of inbound and outbound vessel stop hours as performance indicators, revealing a complete shutdown of vessel traffic during the hurricane and the full recovery within two weeks. From the network perspective, Huang et al., (2024) examined the effects of Typhoon Ma-on on East and Southeast Asia ports through constructing the container shipping networks and highlighting that a 500-km impact range on vessel flows in this TC event. Collectively, the studies of individual TCs offered some detailed insights into how ports have responded to past disruptions.

However, these case studies lacked the generalizability of findings across TCs and provided limited guidance for broader resilience planning and preparedness efforts (Li, 2023). To gain a more comprehensive understanding of port disruption under TCs, analyses covering multiple events and diverse port contexts are essential. (Verschuur et al., 2020) analyzed 141 port disruption events worldwide across 74 ports, but their analysis includes only cases with significant port disruptions or shutdowns. Ports that were affected but continued operating at reduced capacity were excluded, limiting the ability to capture a full spectrum of disruption impacts. Li et al., (2025) examined global port vulnerability under multiple hazards (e.g., TCs, earthquakes, and flooding) from 2015 to 2019, but their analysis focused on aggregated impacts across hazard types. Therefore, it is imperative to develop a comprehensive TC-specific dataset covering multiple TC events and capturing the full range of their impacts on ports, including not only significant disruptions but also moderate, minimal, or no effects, to better understand port resilience.

---

[1] To be consistent, we will use the term 'tropical cyclones' since it is the official terms that describe hurricanes, typhoons and cyclones.

Another key challenge in understanding port resilience to TC impacts lies in unobserved heterogeneity, referring to differences across ports that are not captured in the data but still influence outcomes. For example, two ports may have similar infrastructure and TC exposure, yet one may recover more quickly due to stronger leadership or informal coordination. These unobserved factors contribute to variation in outcomes (Malyshkina et al., 2009). If not properly accounted for, they may lead to biased or misleading conclusions, as the effects of these unobserved factors might be incorrectly attributed to observable variables (Anis et al., 2025; Z. Li et al., 2025; Lord et al., 2021).

This limitation can be partially addressed by integrating multisource datasets to capture a broader range of variables, including those related to ports, TCs, meteorological conditions, and their interactions. While existing studies have combined several data sources (Touzinsky et al., 2018; Verschuur et al., 2020), they remain limited in scope, highlighting the need for an integrated dataset with broader coverage and greater detail. On the other hands, advanced modeling approaches are also critical for addressing the observed heterogeneity. Most existing studies rely on traditional statistical models, such as linear regression, count-based models, and their extensions (Balakrishnan et al., 2022; Verschuur et al., 2023, 2020; Zhou et al., 2020). These models typically treat parameters as fixed across all observations, which limits their ability to capture variation across ports and events. As a result, they often oversimplify complex relationships and fail to reflect the context-specific dynamics critical to understanding port resilience. To address this limitation, advanced modeling approaches are needed. In particular, statistical models with random parameters are essential for capturing unobserved heterogeneity. Unlike fixed-parameter models, random parameter models assume that certain coefficients follow probability distributions, allowing the influence of variables to vary across observations. This flexibility enables more accurate representation of real-world differences in how ports respond to tropical cyclone impacts.

To holistically address the existing limitations, this study begins by constructing the *U.S. Tropical Cyclone Impact on Ports (CyPort) dataset*, which is the first structured and openly available dataset focused specifically on port resilience under TC. CyPort integrates multiple data sources, including AIS vessel trajectory data, tropical cyclone records, and detailed port and meteorological characteristics from 2015 to 2023. Spanning 90 named TCs and 145 U.S. principal ports[2], the dataset enables consistent impact identification and supports scalable, long-term empirical analysis.

To generate more reliable insights into port resilience under tropical cyclones, this study applies a Random Parameter Negative Binomial Lindley (RPNB–Lindley) model (Lord et al., 2021, 2005), which addresses two key challenges: data imbalance and unobserved heterogeneity in port responses. Analysis of the CyPort dataset shows that disruption outcomes are highly imbalanced, with most ports experiencing little to no disruption, while a small number face severe impacts from a given TC (Balakrishnan et al., 2022). Additionally, ports with similar observed characteristics may respond differently due to unobserved factors. The RPNB–Lindley model accommodates these complexities, allowing for more accurate estimation of disruption and recovery patterns and supporting a deeper and reliable understanding of port resilience.

---

[2] Principal Ports refer to the top 150 U.S. ports ranked by total annual tonnage, representing key national gateways (USDOT BTS, 2022). Due to the scarcity and limited access of AIS data in Alaska, 5 ports in Alaska are excluded resulting in a final set of 145 Principal Ports.

## 2. CyPort Dataset

This section describes the data sources and the process used to construct the CyPort dataset. Section 2.1 outlines the scope of the dataset, including its key components and data sources. Section 2.2 explains the construction pipeline, covering how baseline conditions are defined, cyclone impact periods are identified, and resilience metrics are calculated at both the port and network levels.

### 2.1 Dataset Scope and Data Sources

The CyPort dataset is a comprehensive, multisource dataset designed to analyze disruptions to U.S. ports and freight networks caused by TCs and to capture the complexity of port–cyclone interactions. A port–cyclone interaction is defined as the case where a port falls within the buffered impact zone of a tropical cyclone (Huang et al., 2024). In total, CyPort compiles 1,927 interaction records from 2015 to 2023, encompassing 145 U.S. principal ports and 90 named tropical cyclones.

To support detailed analysis, CyPort integrates four key components: (1) TC-disrupted port and freight network data, (2) normal port and freight network data, (3) cyclone exposure data, and (4) port characteristics (**Fig. 1**). Disruption is quantified by comparing vessel activity during TC events with baseline operation conditions. Specifically, performance is measured using port-level indicators (e.g., daily vessel counts) and network-level indicators (e.g., changes in connectivity and vessel flow across the freight network). Both disrupted and baseline metrics are derived from AIS records of commercial vessels. CyPort also incorporates port attributes, such as infrastructure, local demographics, network centrality, and cyclone exposure features drawn from ASOS and CO-OPS weather observations and storm track archives. In total, the dataset provides over 40 interaction variables from 9 distinct sources and generates 7 resilience metrics that reflect disruption magnitude and recovery patterns, as detailed in **Appendix 1**.

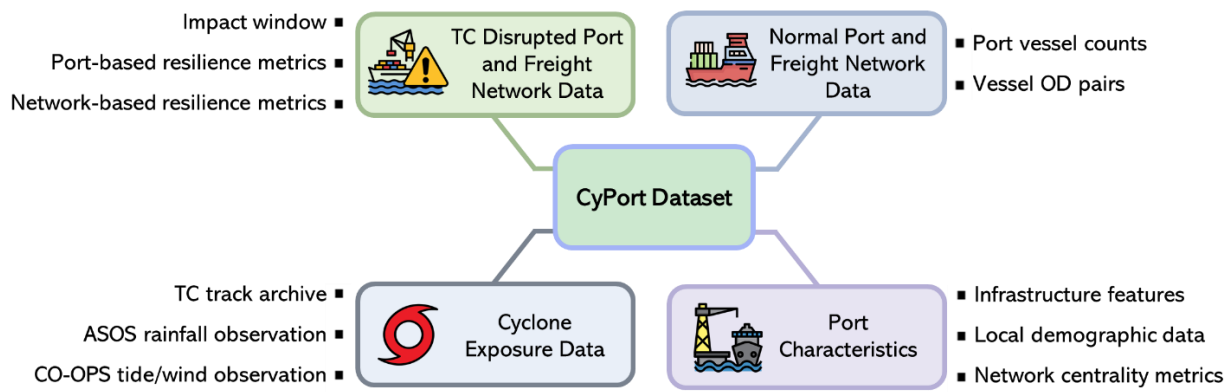

**Fig. 1. CyPort dataset elements and features**

## 2.2 Data Construction Process

The data construction framework consists of three key steps: AIS data cleansing and processing, impact window identification, and cyclone impact evaluation (**Fig. 2**). The process begins with cleaning and aggregating AIS trajectory data to identify vessel visits and movements across ports. This enables the construction of normal performance metrics for both ports and the freight network. Further, baseline performance has been derived to represent a port's normal operations in the absence of TC exposure. Using this baseline, we then detect TC impact windows by comparing vessel activity during cyclone events against historical norms. Finally, the extent of disruption is evaluated by quantifying changes in performance during TC exposure. The framework incorporates both port-level indicators (e.g., vessel volume) and network-level metrics (e.g., node degree) to comprehensively capture disruption. Each step is described in detail in the following subsections. Since the section details the data construction with a focus of port and freight network impact evaluation, details on the integration of cyclone exposure data and port characteristics could be found in **Appendix 1**.

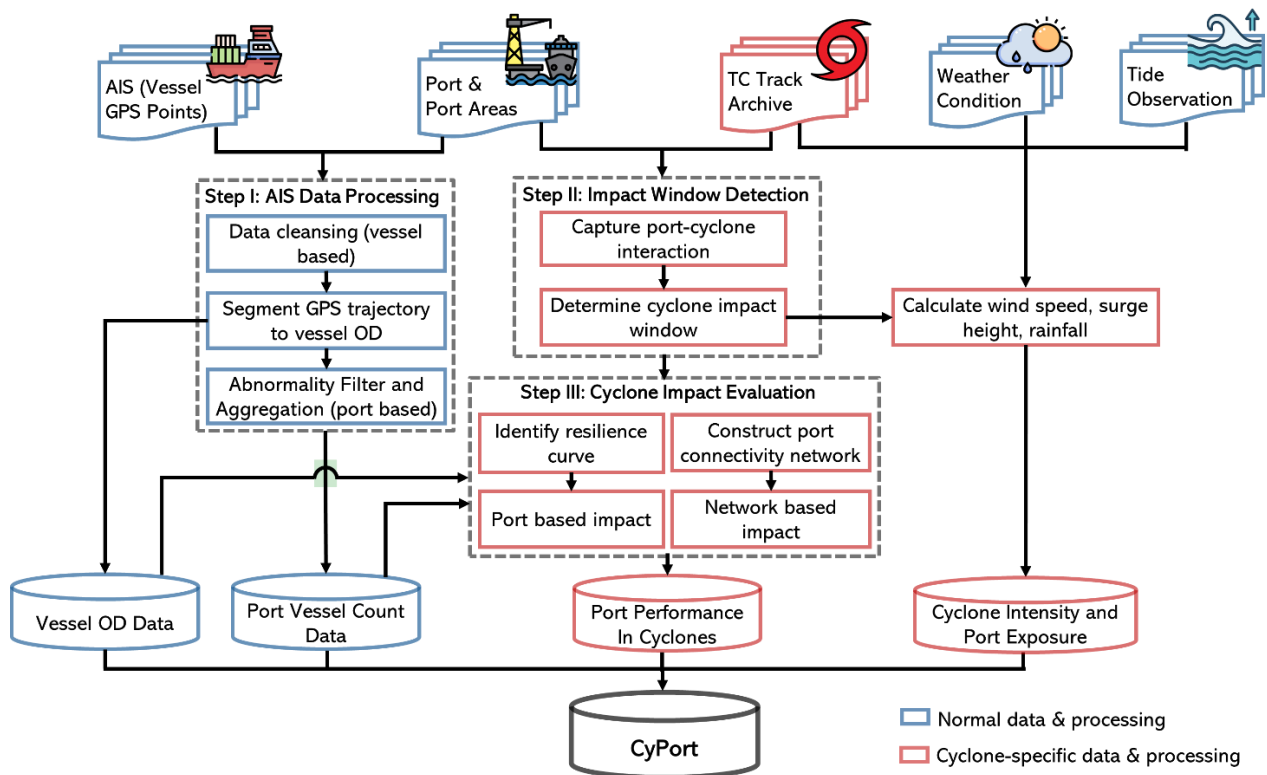

**Fig. 2.** Data construction framework for CyPort dataset

### 2.2.1 Normal port and freight network data

AIS data provides high-frequency records of vessel position, speed, heading, and voyage details every 2 to 10 seconds. Because it captures vessel operations during both normal and disrupted periods, it enables consistent analysis of system performance across conditions (Bläser et al., 2024; Yang et al., 2019). In this study, U.S. AIS data from 2015 to 2023 were obtained, originally collected by the U.S. Coast Guard (Marine Cadastre., 2024). To focus on the freight shipping, we extracted only the trajectories of commercial vessels (i.e. cargo and tanker), while vessels serving other purposes (e.g., passenger, fishing, or pilot operations) were excluded.

We first segmented each vessel's trajectory into a sequence of port visits (**Fig. 3**). Each GPS point was spatially matched to predefined U.S. port boundaries (USAGE, 2021) to determine whether a vessel was entered a port region. Visits lasting less than four hours were excluded to ensure that only actual port calls involving service or cargo operations were captured (Yan et al., 2024). Based on the remaining visits, two datasets were constructed by aggregating port visits: (1) daily vessel counts at each port and (2) origin–destination (OD) pairs representing commercial vessel movements in the freight network.

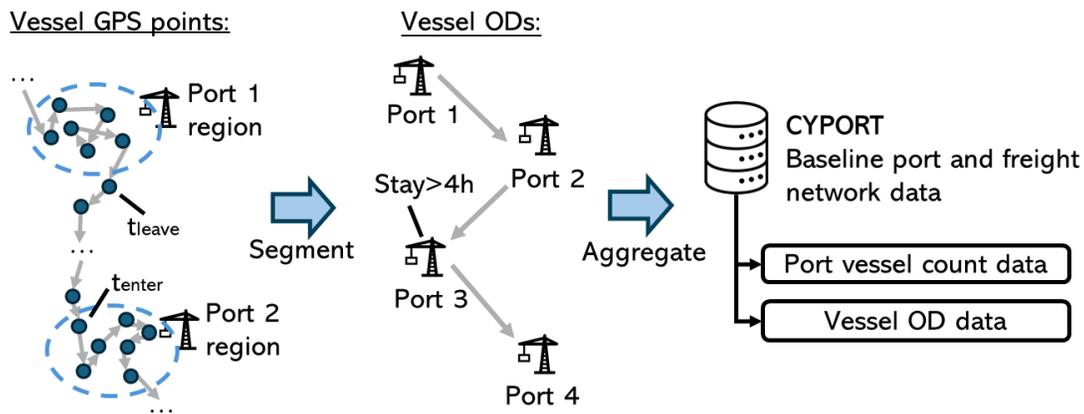

**Fig. 3.** Normal freight system data collection from AIS vessel trajectory

### 2.2.2 Cyclone impact records
#### 2.2.2.1 Cyclone impact window detection

As TCs move inland, they typically exert the most severe impact near their eyes where wind intensity peaks, making ports in this zone particularly vulnerable. Following prior studies, we define the impact zone of a TC as a 500-kilometer buffer surrounding its eye's trajectory (Fang et al., 2022; R. L, 1987). A port-cyclone interaction is thus defined as the event of a port's location falling within this impact zone. The time span during which a port remains in the cyclone's impact zone is considered as the CT's exposure period for the port. However, the actual operational disruption to port's freight activity usually differs from the CT exposure period. For instance, commercial vessels may proactively reroute days before the cyclone's arrival, resulting in an early decline in traffic, and port recovery could continue long after the cyclone has passed, as shown in **Fig. 4(b)**. Therefore, we propose a cyclone impact window detection algorithm to identify the actual impact window.

The first step is to establish a baseline representing a port's normal operations in the absence of TC exposure. This baseline is essential for identifying disruptions, as it provides a reference point to distinguish between typical fluctuations in vessel traffic and actual cyclone-induced impacts. Without such a reference, observed changes during a TC event could be misinterpreted as normal variability. To construct this baseline, we masked the time window around each TC event and used the Prophet time series model to estimate the expected daily vessel counts during the disruption period (**Fig. 3(a)**). The masked window spans from 10 days before to 10 days after the TC exposure, covering both anticipatory and recovery phases (United Nations, 2022). Prophet, developed by Facebook's Core Data Science team, is well suited for this task due to its ability to capture multiple seasonal patterns and incorporate external regressors (Taylor and Letham, 2017). The model takes as input the historical daily vessel counts outside the masked window and produces as output the predicted vessel counts for the masked period along with 95% confidence intervals (CIs). These confidence intervals allow us to detect statistically significant deviations during the TC window. Ports with greater fluctuations in daily vessel count are associated with wider confidence intervals, allowing the algorithm to adapt to varying levels of operational variability and maintain robustness in outlier detection.

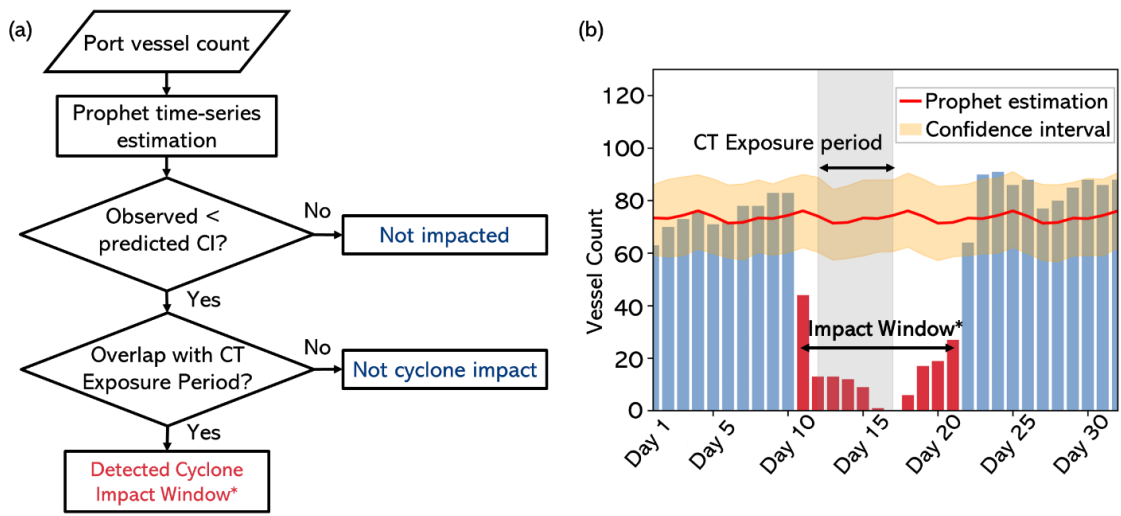

**Fig. 4.** Cyclone impact window detection: (a) detection algorithm; (b) illustration of detected impact window.

Periods in which the observed vessel count fall below the lower bound of the CI are identified as outlier period (Li, 2023), indicating port disruptions. Then, we examine outlier periods' overlap with the CT exposure period. If an overlap is found, the underperformance is attributed to the TC, and the corresponding outlier period is identified as the impact window for that specific cyclone–port interaction, as illustrated in **Fig. 4 (b)**. In many cases, however, no impact window is detected, indicating that the port maintained normal vessel traffic during the TC. The detected impact window serves as the study period of impact evaluation. Examples demonstrating the effectiveness of the proposed impact window detection algorithm are provided in **Appendix. 2**.

#### 2.2.2.2 Port-based impact evaluation

Empirical studies have shown that port operations affected by cyclones exhibits clear resilience curve patterns, e.g. preparation, closure, and recovery phases (Touzinsky et al., 2018). Resilience curve is a well-established concept to communicate quantitative and qualitative aspects of system behavior and resilience (Poulin and Kane, 2021; Zhou et al., 2021). To utilize this tool and evaluate port performance, we apply the time series of vessel counts for each port, denoted as $\{c(t)|t = 1, 2, ..., T\}$, as demonstrated in **Fig. 5**.

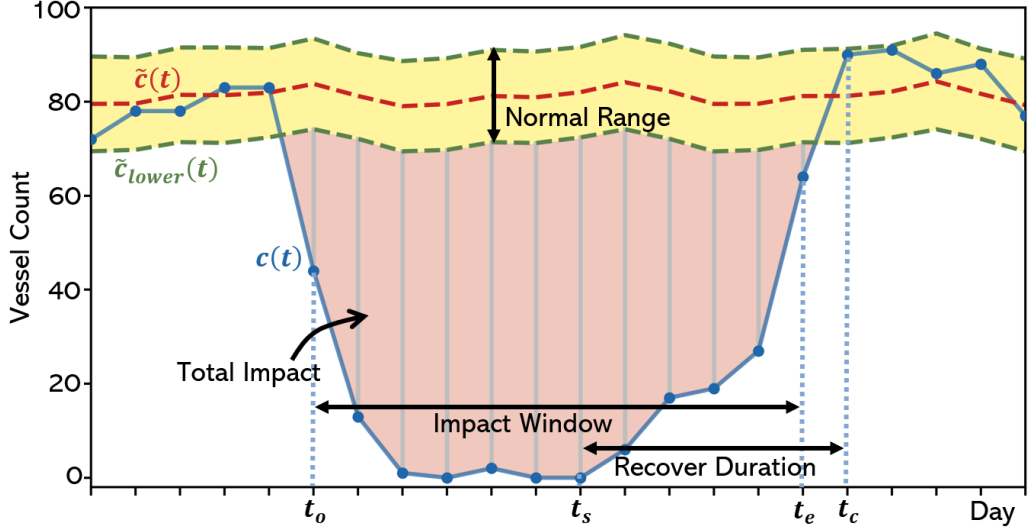

**Fig. 5.** Illustration of port resilience curve

Following previous Prophet estimation, the forecast $\tilde{c}(t)$ and its 95% CI has defined the baseline port performance, which is the normal range of vessel count in **Fig. 4**. The detected impact window $[t_o, t_e]$ is the period during which the port performance is lower than the normal operation and impacted by TC. The key time stamps of port resilience curve are therefore defined:

- $t_o$ (Initial disruption): The day when vessel count performance first drops below the normal range, equal to $t_s$, indicating the onset of abnormal operations.
- $t_s$ (Start of recovery): The point when stable degraded performance ends and vessel counts begin a sustained upward trend. It is identified as the earliest time point, moving backward from the end of the impact window ($t_e$), where the daily change in vessel count ($c(t+1) - c(t)$) becomes and remains positive. This backward-tracing method ensures that temporary fluctuations during the degraded phase are not misclassified as the start of recovery.
- $t_c$ (End of disruption): The point when vessel counts return to the normal operational range within the confidence interval, equal to $t_e + 1$.

Building upon the defined resilience curve, two primary metrics are introduced to quantify and compare resilience patterns across disruptions:

**Total Impact:** The concept of total impact was originally proposed in Bruneau et al.'s foundational work on resilience (Bruneau et al., 2003), summing up the performance loss over a disruption period. The total impact $\mathcal{A}$ is computed as the area under the curve (AUC) between the lowest normal performance and the actual vessel count during the period (Poulin and Kane, 2021):

$$\mathcal{A} = \sum_{t=t_0}^{t_c} 1 - \frac{c(t)}{\tilde{c}_{lower}(t)} \qquad (1)$$

Where $c(t)$ is the vessel count of the port during this disruption at day $t$, $\tilde{c}_{lower}(t)$ is the lower bound of the normal range at the same day. The AUC quantifies the total performance loss, expressed as the equivalent number of days of vessel activity lost. Further, an unnormalized AUC has also been included in the dataset, indicating how many vessel-days are being impacted.

**Recover Duration:** The Recover Duration metric assesses the days required for ports to return to normal operation in terms of vessel counts. It is defined as the day count between start of recovery and end of disruption:

$$T = |t_c - t_s| \qquad (2)$$

Note that if no impact window has been detected, the port maintained normal or even elevated vessel count ($c(t) \geq \tilde{c}_{lower}(t)$). Consequently, the total impact and recover duration are all zero according to equation (1) and (2), meaning no impact found in this port-cyclone interaction.

### 2.2.2.3 Network-based impact evaluation

Networks offer another valuable dimension for analyzing freight systems, providing insights into how tropical cyclones reshaped the vessel traffic and port connectivity (Tafur and Padgett, 2024). Based on the vessel OD dataset, we construct graphs to describe the structure of freight network. Considering the typical long duration of vessel voyage (BTS and FHWA, 2012), we aggregate vessel OD on a weekly basis to better capture temporal patterns. At week $w$, the connectivity of ports and freight network is represented by an undirected graph $\mathcal{G}^w = \{\mathcal{V}, \mathcal{E}^w\}$, where set $\mathcal{V} = \{v_1, v_2, \ldots, v_N\}$ denotes the set of vertex (i.e. port) with $N$=145, and $\mathcal{E}^w \subseteq \mathcal{V} \times \mathcal{V}$ captures port connectivity relationships. Specifically, an undirected edge $\{v_i, v_j\} \in \mathcal{E}^w$ exists if $v_i$ and $v_j$ are visited by a same commercial vessel during week $w$. Note that if we used weighted graphs with vessel counts as edge weights, the degree difference to be calculate would reflect vessel volume reduction and closely match the vessel loss already measured by the resilience curve. To avoid repeating the same information and to clearly show how connections between ports change, we keep the network unweighted. As illustrated in **Fig. 6**, port connectivity can be substantially affected during the week of TC impact ($\overline{w}$), compared to unaffected weeks.

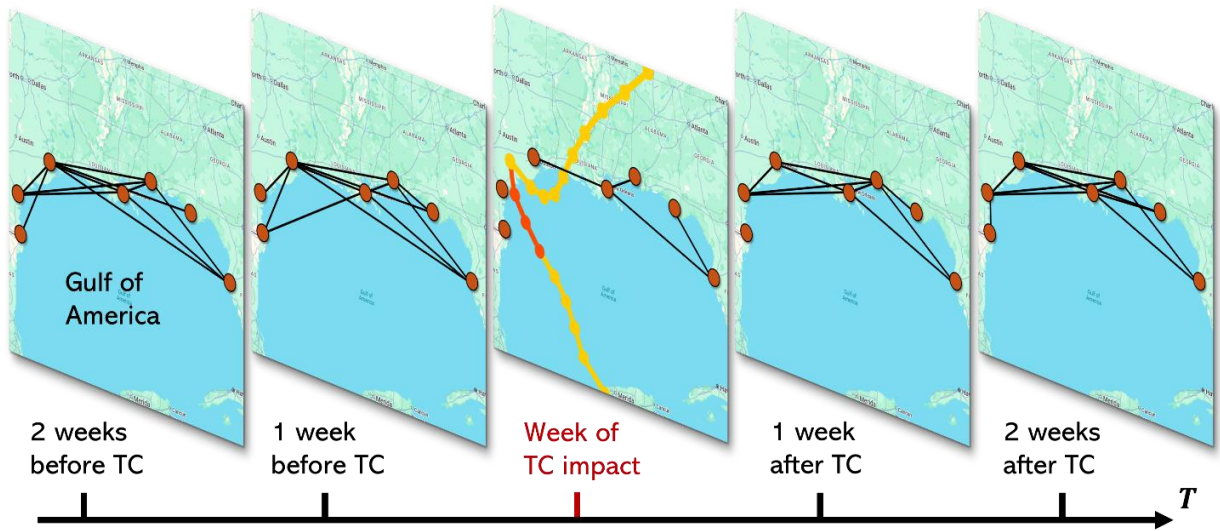

**Fig. 6.** Temporal evolution of the network from two weeks before to two weeks after Hurricane Harvey (2017). The selected nodes (i.e. ports) are located around the Gulf of America to highlight the impact region during the cyclone week.

To evaluate port connectivity that is impacted by tropical cyclones, we compare the same node's connections across weeks: the network of the TC-impacted week $G^{\bar{w}}$ and networks during recent normal weeks, e.g. $G^{\bar{w}-2}$. As adopted from previous studies (Huang et al., 2024; Peng et al., 2018), we apply degree as simple and direct graph connectivity measurement. The degree sequence of port node $n$ during the weeks around the tropical cyclone week $\bar{w}$ is given by $\{d(v_n)_{\bar{w}-M}, \dots, d(v_n)_{\bar{w}}, \dots, d(v_n)_{\bar{w}+M}\}$, where $M$ is the selected number of recent unaffected weeks before and after TC to represent the recent normal performances. In this study, we adopted $M = 4$, which considers about 1 month before and after the TC week to get baseline performance (Chen et al., 2016). Therefore, port average degree and the **degree difference** $\Delta d$ of port n caused by the tropical cyclone is given by:

$$d(v_n) = \frac{1}{2M} \sum_{m=1}^{M} d(v_n)_{\bar{w}-m} + d(v_n)_{\bar{w}+m} \tag{3}$$

$$\Delta d = |d(v_n) - d(v_n)_{\bar{w}}| \tag{4}$$

The degree difference estimates the loss of network connectivity due to TC. Besides quantifying disruptions, we also compute graph-theoretic centrality metrics through averaging across recent normal weeks to denote the port's role in freight network:

$$C_{cls}(v_n) = \frac{1}{2M} \sum_{m=1}^{M} \frac{N-1}{\sum_{v_j \neq v_n} \delta_{\bar{w}-m}(v_n, v_j)} + \frac{N-1}{\sum_{v_j \neq v_n} \delta_{\bar{w}+m}(v_n, v_j)} \tag{5}$$

$$C_{btw}(v_n) = \frac{1}{M(N-1)(N-2)} \sum_{m=1}^{M} \sum_{v_i \neq v_n \neq v_j} \frac{\sigma_{\bar{w}-m}^{ij}(v_n)}{\sigma_{\bar{w}-m}^{ij}} + \frac{\sigma_{\bar{w}+m}^{ij}(v_n)}{\sigma_{\bar{w}+m}^{ij}} \tag{6}$$

Where $\delta_w(v_n, v_j)$ is the shortest path length between node $v_n$ and $v_j$ given week $w$. $\sigma_w^{ij}$ represents the number of shortest paths between any nodes $v_i$ and $v_j$ at week $w$, $\sigma_w^{ij}(v_n)$ is the number of these shortest paths that pass through node $v_n$.

Closeness centrality ($C_{cls}$) measures the inverse of the average shortest path distance from a port to all others in the network, characterizing its ability to efficiently reach all regions, and is widely interpreted as an indicator of transshipment center in maritime logistics (Huang et al., 2024). Betweenness centrality ($C_{btw}$), which quantifies how often a port lies on the shortest paths between other port pairs, captures the port's role as a routing intermediary, enabling path flexibility and network-wide vessel volume redistribution in case of disruptions (Perez and Germon, 2016). Degree centrality, same as node's degree ($d(v_n)$), directly reflects the number of direct port-to-port connections a port maintains and is commonly used to identify major hub ports with high physical connections (Govindan et al., 2017; Newman, 2018). These centrality measures collectively provide insight into the functional roles of ports in the freight network and support a structural explanation of how network position contributes to disruption patterns.

## 3. Statistical Modeling of Port Resilience

In the previous section, we introduced the construction of the CyPort dataset. Building on this dataset, we apply an advanced statistical model to analyze port disruptions, addressing key challenges such as excess zero observations inherent in the data and unobserved heterogeneity in port resilience. This approach provides more reliable insights across multiple TC events, moving beyond the limitations of traditional case studies.

### 3.1 Excess Zero Observations of TC Impact on Port

This study selects three key resilience indicators introduced earlier as dependent variables: total impact, recovery duration, and degree difference. Total impact is defined as the number of equivalent days lost in vessel activity due to the tropical cyclone, serving as a proxy for disruption severity. Recovery duration captures the time required for a port's operational performance to return to the normal level, reflecting the speed of recovery. Degree difference measures the extent to which its connectivity within the freight network is affected. In this study, all of them are modeled as count variable (non-negative integer). Recovery duration is inherently a count of days. Degree difference and total impact, while originally non-integral, are rounded to the nearest integer to reflect the discrete nature of port degree changes and impacted days of volume, respectively. This approach is consistent with prior studies that treat discretized resilience outcomes as count responses for regression modeling (Balakrishnan et al., 2022).

In low-traffic ports, reductions in vessel activity are often difficult to attribute correctly to cyclone impacts over daily fluctuations, introducing bias into impact assessments (Verschuur et al., 2020). Therefore, ports with an average daily vessel count below five (e.g., Port of Grays Harbor, Port of Terrebonne) were excluded. This filtering step helps mitigate noise and ensures more reliable statistical inference. As a result, 1,093 out of the original 1,927 port–cyclone interaction records were retained for analysis.

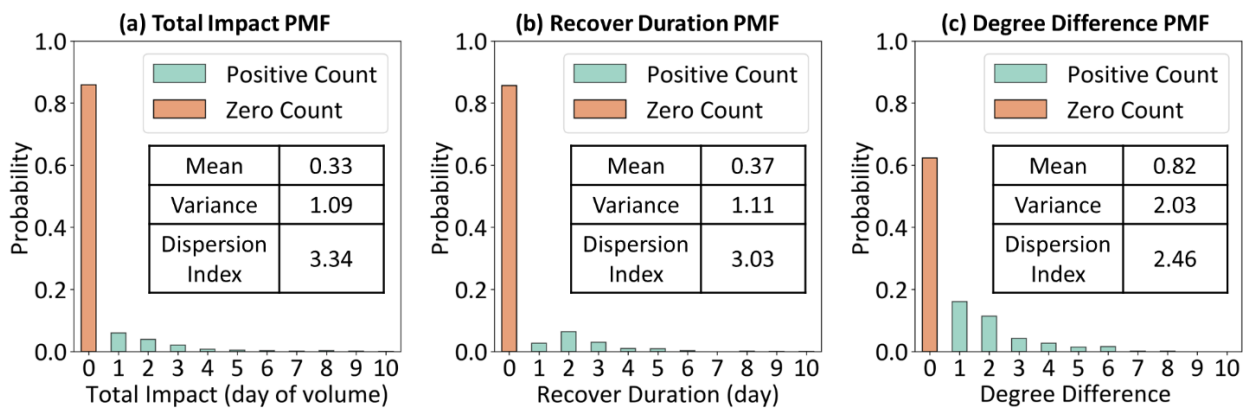

**Fig. 7.** Probability Mass Functions (PMFs) and Dispersion Characteristics of resilience metrics: (a) Total Impact, (b) Recovery Duration, and (c) Degree Difference.

The revealed resilience metric, however, demonstrates clear zero-inflated patterns: during tropical cyclone events, although ports are frequently exposed to disruptive conditions, a large proportion exhibits no significant impact and continue normal operations (**Fig. 7**). This phenomenon aligns with findings from prior port resilience studies (Balakrishnan et al., 2022), which highlight the prevalence of excessive zero outcomes in disruption datasets. In addition to zero-inflation, the data also exhibits over-dispersion, as evidenced by dispersion index (the variance-to-mean ratio) values exceeding 1. These distributional characteristics violate the assumptions of conventional Poisson models, which require equal mean and variance, and justify the use of more flexible count data models.

### 3.2 Random Parameters Negative Binomial-Lindley (RPNB-Lindley) Model

Regression analysis has been widely adopted in resilience analysis to study the relationships between hazard impact and their contributing factors (Han, 2018). However, modeling port disruption during tropical cyclones presents several challenges. The observed count data are marked by an excess number of zeros (i.e., datasets with a large number of zero observations) and over-dispersion data pattern. Furthermore, the effects of cyclone exposure vary significantly across ports due to differences in physical infrastructure, location, and institutional resilience. The RPNB–Lindley model addresses the needs through two extensions on traditional negative binomial models. First, it introduces a Lindley-distributed random effect into the expected count structure to model over-dispersion and structural zeros (Rusli et al., 2018). The NB-L offers sound theoretical grounds in which the long-term mean of the distribution is never zero (Geedipally et al., 2012; Lord et al., 2005). Second, the random parameters allow selected regression coefficients to vary randomly across observations, enabling the model to account for unobserved heterogeneity across disruptions (Khan et al., 2023).

The core model assumes the observed impact $Y_k$ (i.e. total impact, recover duration and degree difference) of the $k$ th port–cyclone interaction follows a Negative Binomial (NB) distribution controlled by $\theta_k$ as the expected count and $\varphi$ as a dispersion parameter. Its distribution and probability mass function are then defined as (Shaon et al., 2018):

$$Y_k \sim NB(\theta_k, \varphi) \tag{7}$$

$$P(Y_k \mid \theta_k, \phi) = \frac{\Gamma(Y_k + \phi)}{Y_k!\, \Gamma(\phi)} \left(\frac{\phi}{\phi + \theta_k}\right)^{\phi} \left(\frac{\theta_k}{\phi + \theta_k}\right)^{Y_k} \tag{8}$$

To consider the excess zero observations data pattern, the mean parameter $\theta_k$ is modeled as a product of a deterministic component $\lambda_k$ and a Lindley-distributed random effect $\delta_k$. In this model, the $\delta_k$ is implemented through a two-step hierarchical structure: a Bernoulli trial determines the gamma shape parameter, and $\delta_k$ is then drawn from a gamma distribution:

$$\theta_k = \lambda_k \delta_k \tag{9}$$

$$\delta_k \sim \text{Gamma}(1 + z_k, \psi) \tag{10}$$

$$z_k \sim \text{Bernoulli}\left(\frac{\psi}{1 + \psi}\right) \tag{11}$$

This formulation allows the shape of the distribution to vary across observations and approximates a Lindley distribution. The hyperparameter $\psi$ governs both the degree of zero-inflation and the fat tail behavior. Together, equations (9) – (11) introduce flexible dispersion patterns and structural zeros into the model.

With the Lindley-distributed random effect $\delta_k$ capturing the zero-inflation data pattern, the deterministic mean component $\lambda_k$ captures the influence of observable factors (e.g., cyclone intensity, wind speed and port location) and is specified via a log-linear function:

$$\lambda_k = exp\left(\beta_0 + \sum_{l=1}^{q} \beta_l^k x_{kl}\right) \tag{12}$$

Here $x_{kl}$ denotes the value of the lth variable for the $k$th port cyclone interaction, and $\beta_l^k$ is the corresponding regression coefficient.

To incorporate the random parameter capturing unobserved heterogeneity across port disruptions, the subset of coefficients $\{\beta_l^k\}$ are allowed to vary randomly across observations, capturing latent differences in how each explanatory variable influences the outcome. This is achieved by decomposing the random coefficient as:

$$\beta_l^k = \beta_l + v_l^k \tag{13}$$

$$v_l^k \sim \mathcal{N}(0, \sigma_l^2) \tag{14}$$

Where $\beta_l$ is the global mean effect of $l$th variable, and $v_l^k$ is a zero-mean gaussian random variable capturing heterogeneities of port impact.

Therefore, the likelihood function for the RPNB–Lindley model can then be expressed as integral of random variables (Oviedo-Trespalacios et al., 2020):

$$P(Y_k = y_k) = \int \frac{\Gamma(y_k + \varphi)}{y_k!\,\Gamma(\varphi)} \left(\frac{\varphi}{\varphi + \theta_k}\right)^{\varphi} \left(\frac{\theta_k}{\varphi + \theta_k}\right)^{y_k} g(v_l^k)\, dv_l^k \tag{15}$$

Where $g(v_l^k)$ is the standard normal density function. This integral formulation reflects the contribution of both fixed and random effects to the outcome distribution. During model estimation, the parameters $\beta_l$, $\sigma_l^2$, $\psi$ and $\phi$ are estimated jointly. The selection of the random parameter variables was based on (Anis et al., 2025).

The model is estimated under a Bayesian hierarchical framework, using Markov Chain Monte Carlo sampling. Bayesian estimation is particularly advantageous in this setting, as it facilitates uncertainty quantification across all levels of the model—fixed effects, random parameters, dispersion, and excess zero observations structure, and yields full posterior distributions for interpretation (Rosen et al., 2022). The data was randomly split into a training set and a test set in an 80:20 ratio. The training set was used for parameter fitting, while the test set was used to evaluate model performance and calculate the marginal effects.

The inclusion of Negative Binomial (NB) and Negative Binomial–Lindley (NB–Lindley) models serves as an essential baseline model for comparison. While the NB model controls for overdispersion, it does not account for structural zeros. The NB–Lindley model addresses excess zeros and additional dispersion via a Lindley-distributed random effect but assumes fixed regression effects. By comparison, the RPNB–Lindley model integrates both excess zero observations and random coefficients, offering enhanced flexibility to model data patterns and unobserved heterogeneity in CT impact. The model fit and predictive performance are evaluated using three measurements: the Deviance Information Criterion (DIC) for overall model adequacy with complexity penalization, and the Mean Absolute Error (MAE) and Root Mean Square Error (RMSE) for assessing the model's ability to reproduce observed counts.

### 3.3 Independent Variable Selection

The Independent Variables are categorized into three groups: TC exposure, port characteristics, and port centrality measures, as summarized in **Table 1**. TC exposure variables capture the cyclone's intensity category, the minimum distance between the cyclone center and the port, and local weather conditions experienced at the port. Port characteristics, e.g. geographic location, population are evaluated since it represents the location as well as local resourcefulness against TC (U.S. Census Bureau., 2021). Port centrality measures has been introduced before reflecting the structural importance of each port in recent weeks' networks.

**Table 1.** Description and data summary of Independent Variables

| | Variable | Description | Mean | St.D. | Min. | Max. |
|---|---|---|---|---|---|---|
| TC Exposure | Wind Speed | Maximum sustained wind speed (knots) at port during the cyclone | 22.40 | 9.58 | 7.78 | 81.23 |
| | Rainfall | Total accumulated rainfall (mm) at port during the cyclone | 8.99 | 26.05 | 0 | 422 |
| | Surge Height | Maximum observed storm surge height (meters) at port during the cyclone | 0.42 | 0.36 | -0.23 | 3.18 |
| | Distance | Closest distance (km) from cyclone center to the port | 242.07 | 133.82 | 1.8 | 494.9 |
| Port Characteristics | East Coast | Indicator for port's location in east coast (0/1); Non-coast serves as reference category | 0.450 | - | 0 | 1 |
| | Pacific Coast | Indicator for port's location in Pacific Coast (0/1) | 0.018 | - | 0 | 1 |
| | Gulf of America | Indicator for port's location in Gulf of America (0/1) | 0.396 | - | 0 | 1 |
| | Ln(Population) | Natural logarithm of port's county population | 12.84 | 1.32 | 9.91 | 16.12 |
| | Tract Workforce Factor | Disadvantaged workforce indicator in port's census tract (0/1) | 0.29 | - | 0 | 1 |
| | Tract Poverty Rate | Share of tract population below federal poverty line | 0.51 | 0.35 | 0 | 0.99 |
| | Transportation Insecurity | Indicator for limited or unreliable access to transportation (0/1) | 0.52 | - | 0 | 1 |
| | Transportation Access | Indicator for adequate availability of transportation (0/1) | 0.45 | - | 0 | 1 |
|  | Docks | Number of operational docks at the port | 32.40 | 45.68 | 1 | 238 |
|  | Railway | Length of railway within 3km buffer zone of port (m). | 69.07 | 72.37 | 0 | 328.4 |
|  | Highway | Length of highway within 3km buffer zone of port (m) | 52.03 | 38.32 | 0 | 156.0 |
| Network Centrality measures | Degree Centrality | Port degree centrality during normal weeks | 7.06 | 3.99 | 0 | 20 |
|  | Closeness Centrality | Port closeness centrality during normal weeks | 0.18 | 0.046 | 0 | 0.274 |
|  | Betweenness Centrality | Port betweenness centrality during normal weeks | 0.009 | 0.011 | 0 | 0.064 |

All variables are selected from the processed CyPort dataset. To ensure optimal predictor selection and address multicollinearity, a two-step variable selection process was conducted. Variance Inflation Factor (VIF) test was first applied, and variables with VIF > 5 were excluded to mitigate multicollinearity issues (Marcoulides and Raykov, 2019). Through this test, Transportation Insecurity and Transportation Access are excluded from analysis, due to high collinearity with other variables. Next, stepwise variable selection based on the smallest deviance information criterion (DIC) was performed to identify the most relevant explanatory variables, optimizing model performance (Hu et al., 2021). It is worth mentioning that the VIF test and stepwise selection does not yield the same optimal variable set for different resilience metrics. Thus, some variables are excluded from one model but still retain in other models, to ensure the highest DIC performance for each model.

### 3.4 Marginal Effect

To gain deeper insights into the influence of the contributing factors on port impact outcomes, the marginal effects of statistically significant variables were computed from the test set of data. For continuous explanatory variables, marginal effects quantify the change in the expected value of the impact metric associated with a one-unit increase in the variable, holding all other explanatory variables constant. For discrete or categorical variables, marginal effects represent the change in the expected outcome when the variable shifts from the reference category to a specific category. For each independent variable $x_l$, the marginal effect at observation $k$ considering parameter uncertainty and the average marginal effect (AME) across observations are calculated as follows (Washington et al., 2020):

$$\frac{\partial E(\lambda_k)}{\partial x_{kl}} = E_{\boldsymbol{\beta}_k}[exp(\boldsymbol{\beta}_k^\top \boldsymbol{x}_k) \cdot \beta_{kl}] \tag{16}$$

$$AME_l = \frac{1}{K} \sum_{k=1}^{K} \frac{\partial E(\lambda_k)}{\partial x_{kl}} \tag{17}$$

Where $E(\lambda_k)$ denotes the expected value of the port impact metric (e.g., total impact, recovery duration, or degree difference) for observation $k$, and $x_{kl}$ is the value of the $l$th independent variable. The expectation operator in Equation (14) reflects integration over the distribution of random parameters, computed using Halton draws. Incorporating random parameters into the model allows the effect of each variable to vary across observations, capturing unobserved heterogeneity in behavioral responses or contextual conditions (Khan et al., 2023). The marginal effect represents the change in the expected impact count associated with a one-unit increase in a continuous variable, or a change from 0 to 1 for an indicator variable, while holding all other variables constant (Norton et al., 2019).

## 4. Results and Discussion
### 4.1 Model Performance Evaluation

The model estimation results are summarized in **Table 2**. Under the DIC, MAE, and RMSE evaluations, the RPNB-Lindley Model consistently outperformed the others. Its superior performance can be attributed to the Lindley distribution, which effectively captures the dispersion within the data, and the incorporation of random parameters, which account for unobserved heterogeneity. Therefore, we focus the discussion on the RPNB-Lindley Model in subsequent analyses.

**Table 2.** Performance Comparison of Models for different impact dimensions

|  | DIC | MAE | RMSE |
|---|---|---|---|
| *Total Impact (the normalized vessel volume loss in equivalent days)* | | | |
| NB Model | 1799 | 0.727 | 1.163 |
| NB-Lindley Model | 1749 | 0.554 | 0.837 |
| RPNB-Lindley Model | **1735** | **0.533** | **0.804** |
| *Recovery Duration (Number of days the port took for recovery)* | | | |
| NB Model | 1961 | 0.844 | 1.197 |
| NB-Lindley Model | 1943 | 0.705 | 1.048 |
| RPNB-Lindley Model | **1938** | **0.675** | **0.982** |
| *Degree Difference (Cyclone week reduction in port's node degree)* | | | |
| NB Model | 2634 | 1.073 | 1.407 |
| NB-Lindley Model | 2542 | 0.664 | 0.735 |
| RPNB-Lindley Model | **2499** | **0.649** | **0.698** |

As demonstrated in **Table 3**, the specific coefficient estimates of the model are presented, including their mean, standard deviation (SD), and the lower and upper limits (LL, UL) of the 95% Bayesian Credible Interval (BCI). Because only variables that maximize the DIC are retained, insignificant factors are excluded from the final models. For instance, within the TC Exposure Variables, SSHS Category 3 is only significant in both degree difference and total impact yet not significance (under the 95% interval) for recovery duration and total impact.

Among the TC exposure variables, all port-side meteorology observations, including wind speed, storm surge height, and rainfall, are found to significantly influence port resilience outcomes. However, storm surge height does not significantly affect recovery duration. This finding may be attributed to the nature of storm surge, which causes immediate and severe physical disruptions, thereby contributing to greater total impact and changes in network connectivity. While larger total impacts are generally associated with longer recovery times, our findings suggest that recovery duration is more directly influenced by wind-related physical damage or delays caused by operational and regulatory constraints during the recovery process (United States Coast Guard, 2024). The distance from the cyclone's eye is negatively associated with both total impact and recovery duration, suggesting that ports located closer to the storm core suffer greater disruption and require more time to return to normal functioning. However, distance does not significantly explain the observed changes in network connectivity, as measured by degree difference. One possible explanation is that large tropical cyclones tend to disrupt regional shipping activities over wide areas, making proximity less influential in determining the extent of connectivity loss. Finally, cyclone intensity, as classified by the Saffir–Simpson Hurricane Scale (SSHS), reveals important thresholds of disruption. Detailed discussions are provided in section 4.2.1.

**Table 3.** Posterior estimates of RPNB-Lindley model parameters for different impact dimensions

| Variables | Total Impact | | | Recovery Duration | | | Degree Difference | | |
|---|---|---|---|---|---|---|---|---|---|
| | Mean (SD) | 95% BCI LL | 95% BCI UL | Mean (SD) | 95% BCI LL | 95% BCI UL | Mean (SD) | 95% BCI LL | 95% BCI UL |
| **TC Exposure Variables** | | | | | | | | | |
| SSHS Category 3 (Base: SSHS Category TS) | - | - | - | - | - | - | 0.308 (0.127) | 0.060 | 0.558 |
| SSHS Category 4 | 0.836 (0.225) | 0.397 | 1.280 | 0.818 (0.232) | 0.367 | 1.281 | 0.522 (0.144) | 0.237 | 0.806 |
| SSHS Category 5 | 1.014 (0.257) | 0.513 | 1.523 | 0.886 (0.273) | 0.357 | 1.426 | 0.617 (0.165) | 0.293 | 0.944 |
| Wind Speed | 0.037 (0.016) | 0.005 | 0.068 | 0.026 (0.011) | 0.005 | 0.047 | 0.023 (0.011) | 0.002 | 0.045 |
| Rainfall | 0.007 (0.002) | 0.002 | 0.012 | 0.006 (0.003) | 0.001 | 0.011 | 0.002 (0.001) | 0.0006 | 0.005 |
| Surge Height | 0.213 (0.104) | 0.002 | 0.419 | - | - | - | 0.340 (0.154) | 0.037 | 0.650 |
| Distance | -0.002 (0.0006) | -0.003 | -0.001 | -0.002 (0.0006) | -0.003 | -0.0004 | - | - | - |
| **Port Characteristics and Centrality Measures** | | | | | | | | | |
| Gulf of America (Base: Non-coast ports) | 1.017 (0.326) | 0.380 | 1.261 | 0.830 (0.314) | 0.216 | 1.445 | 0.865 (0.378) | 0.124 | 1.635 |
| East Coast | 1.133 (0.335) | 0.470 | 1.667 | 0.909 (0.329) | 0.269 | 1.565 | 0.420 (0.138) | 0.144 | 0.691 |
| Pacific Coast | 1.710 (0.674) | 0.379 | 1.797 | 1.465 (0.701) | 0.077 | 2.845 | 0.254 (0.122) | 0.015 | 0.493 |
| Degree Centrality | - | - | - | - | - | - | 0.050 (0.024) | 0.002 | 0.098 |
| Betweenness Centrality | - | - | - | - | - | - | -6.570 (3.27) | -12.98 | -0.16 |
| Closeness centrality | 7.636 (3.614) | 0.304 | 5.305 | 5.694 (2.568) | 0.710 | 10.785 | 8.998 (1.837) | 5.443 | 12.629 |
| Ln(Population) | -0.126 (0.064) | -0.253 | -0.000 | - | - | - | 0.090 (0.044) | 0.011 | 0.190 |
| Tract Poverty Rate | -0.011 (0.003) | -0.017 | -0.003 | -0.009 (0.003) | -0.017 | -0.003 | - | - | - |
| Tract Workforce Factor | -0.080 (0.022) | -0.126 | -0.036 | -0.008 (0.002) | -0.013 | -0.003 | - | - | - |
| **Standard Deviation of Random Parameters** | | | | | | | | | |
| Wind Speed | 0.127 (0.074) | 0.048 | 0.226 | 0.048 (0.008) | 0.033 | 0.063 | 0.037 (0.005) | 0.028 | 0.048 |
| Surge Height | 0.067 (0.022) | 0.042 | 0.111 | 0.056 (0.009) | 0.038 | 0.073 | 0.311 (0.004) | 0.024 | 0.040 |
| Rainfall | 0.727 (0.286) | 0.394 | 1.166 | - | - | - | - | - | - |
| Distance | 0.018 (0.003) | 0.013 | 0.023 | 0.015 (0.001) | 0.013 | 0.017 | - | - | - |
| **Model Estimation Parameters** | | | | | | | | | |
| Constant | -1.774 (0.269) | -2.218 | -1.242 | -3.496 (1.692) | -6.349 | -0.712 | -6.285 (1.246) | 8.393 | 3.802 |
| Dispersion Parameter | 1.013 (0.283) | 0.617 | 1.604 | 0.708 (0.183) | 0.442 | 1.148 | 4.477 (1.451) | 2.404 | 8.025 |
| Lindley Parameter | 1.727 (0.785) | 0.205 | 3.248 | 1.389 (0.350) | 0.695 | 0.209 | 0.308 (1.485) | 0.033 | 0.587 |

Note:
[1] "–" indicates parameter insignificant and therefore removed in the final models
[2] Model performance measures (DIC, MAE, RMSE) are shown in Table 4 for RPNB-Lindley models.

Turning to Port Characteristics and Centrality Measures, the coefficient for ln(population) suggests that ports situated in counties with larger populations tend to be more robust (i.e., they experience lower total impact), potentially due to resource availability, disaster preparedness, or more comprehensive infrastructure. Nonetheless, a slightly larger coefficient for degree difference indicates that while physical disruptions may be minimized, extensive connectivity networks are more susceptible to short-term disruptions in these areas.

Regarding random parameters, both wind speed and surge height prove significant across all three impact dimensions, reflecting unobserved heterogeneity in how different ports or storm events respond to these hazard variables. Their random effects suggest that certain ports may exhibit heightened or reduced sensitivity, potentially influenced by factors such as infrastructure resilience or storm preparedness strategies (Peterson, 2000; Snow, 1984). By contrast, rainfall emerges as a random parameter only for total impact, possibly due to the localized nature of flood damage during the initial disruption phase, whereas the variability in extended recovery processes or network disruptions tied to rainfall may be relatively modest.

### 4.2 Port Resilience Effect Analysis
#### 4.2.1 TC Exposure

A non-linear relationship is observed between storm intensity and port resilience, indicating that increases in storm severity do not lead to proportionate increases in disruption. Instead, impacts remain relatively moderate under weaker storms (SSHS categories 1 and 2 are not significant) and then rise sharply beyond a certain threshold (SSHS categories 4, 5 are both significant and have significantly larger average marginal effects across all impact matrices). This pattern is consistent with broader findings in the disaster literature, which note that marginal increases in hazard severity can lead to exponentially greater impacts (Cutter et al., 2008). From a systems perspective, this pattern reflects the resilience tipping point, defined by Holling (Holling,

1973), as the threshold beyond which a system experiences disproportionately large or even irreversible changes in response to external stressors.

**Table 4. Marginal effects for TC exposure variables**

| Variables | Average Marginal Effects | | |
| --- | --- | --- | --- |
| | Total Impact | Recovery Duration | Degree Difference |
| SSHS Category 3 (Base: SSHS Category: TS) | - | - | 0.387 |
| SSHS Category 4 | 0.577 | 0.643 | 0.633 |
| SSHS Category 5 | 0.704 | 0.701 | 0.738 |
| Wind Speed | 0.021 | 0.022 | 0.026 |
| Rainfall | 0.004 | 0.005 | 0.003 |
| Surge Height | 0.139 | - | 0.432 |
| Distance | -0.0013 | -0.0014 | - |

Our results identify this tipping point at SSHS Category 4, where impacts begin to escalate dramatically across all resilience dimensions. While hurricanes classified as Category 3 or higher are labeled as "major" storms (NOAA, 2021), Category 3 events in our analysis show less effects compared to higher SSHS levels, exhibiting statistical significance only for degree difference (0.387), with no significant effect on total impact or recovery duration. In contrast, Category 4 storms demonstrate substantially larger marginal effects and increase further at Category 5. This shift reflects a fundamental change in system behavior once Category 4 intensity is reached. According to NOAA (2021), Category 4 hurricanes cause catastrophic damage, including severe structural failures in well-built homes, power outages lasting weeks to months, and infrastructure collapses that render areas uninhabitable. For ports, this translates into widespread operational disruption, extended recovery timelines, and significant losses in freight network connectivity. These findings indicate that while lower-category hurricanes may cause short-term operational disturbances, the onset of Category 4 conditions signifies a resilience tipping point, leading to fundamentally altered system properties characterized by sharply elevated vulnerability and prolonged disruption.

### 4.2.2 Port Characteristics

Port location plays a crucial role in shaping resilience patterns, as coastal ports experience significantly greater disruptions than their non-coastal counterparts. As shown in **Table 5**, ports along the Gulf of America, East Coast, and Pacific Coasts all exhibit positive and significant coefficients across multiple impact dimensions, with Pacific Coast ports demonstrating the highest marginal effect (1.202) for total impact. However, this finding requires careful interpretation, as Pacific Coast ports have historically been exposed to fewer TC events compared to those along the Gulf of America and East Coast (NOAA, 2025). This discrepancy suggests that regional factors—such as storm preparedness measures, historical exposure to extreme weather, and geographic vulnerability—likely influence the magnitude of impact. Ports in the Gulf Coast, for instance, exhibit the highest coefficient for degree difference (0.865), indicating that their network connectivity is particularly vulnerable during storm events. The stronger disruptions observed in these regions highlight how geographic exposure amplifies cyclone impacts, reinforcing the necessity for region-specific resilience planning. While storm exposure remains the primary driver of resilience outcomes, these findings suggest that the broader regional context, such as coastline vulnerability, emergency response capabilities, and historical adaptation efforts, plays a vital role in shaping disruption severity. This strengthens earlier findings that geographic clustering of exposure and infrastructure fragility creates uneven resilience profiles across port systems (Becker et al., 2012).

**Table 5. Marginal effects for Port Characteristics**

| Variables | Average Marginal Effects | | |
| --- | --- | --- | --- |
| | Total Impact | Recovery Duration | Degree Difference |
| **Port Characteristics** | | | |
| Gulf of America (Base: Non-coast ports) | 0.699 | 0.655 | 0.961 |
| East Coast | 0.778 | 0.717 | 0.463 |
| Pacific Coast | 1.202 | 1.154 | 0.281 |
| Ln(Population) | -0.086 | - | 0.112 |
| Tract Poverty Rate | -0.007 | -0.008 | - |
| Tract Workforce Factor | -0.053 | -0.006 | - |

Beyond geographic location, socioeconomic factors contribute to resilience variation, though their effects are more marginal compared to meteorological and locational variables. County population has a negative and significant effect on total impact (-0.126), suggesting that ports situated in more densely populated areas experience lower levels of disruption. This could be attributed to better resource availability, enhanced emergency management systems, or more extensive infrastructure investments. However, population also exhibits a small positive coefficient (0.090) for degree difference, implying that while physical disruptions may be mitigated in high-population areas, their network connectivity may be more sensitive to disruptions, potentially due to increased traffic volumes and dependencies on complex supply chains. Similarly, higher poverty rates and a stronger workforce factor regions are associated with reduced impact, yet these marginal effects are considerably smaller than those linked to TC exposure or port location. These findings emphasize that resilience strategies should integrate both physical and socioeconomic dimensions, but with primary attention directed toward mitigating exposure risks at given coasts and strengthening infrastructure resilience.

### 4.2.3 Network Features

Port connectivity within the weekly freight network is notably influenced by TCs. **Fig. 8** has illustrated three representative cases: one under normal operating conditions and two during significant disruptions caused by TCs impacting different U.S. coasts. We selected three cases, ensuring temporal proximity to minimize bias due to long-term changes in shipping patterns. From the network structure, it is evident that the Great Lakes ports demonstrate strong internal connectivity, yet they exhibit minimal reachability to external regions. This limited accessibility is attributed to the sparse commercial vessel activity inland and the fact that their primary external connection is via the Mississippi River system, particularly through southern Louisiana (USACE, 2000). The giant connected component of network typically comprises ports along the Gulf and

East coasts, whereas Pacific coast ports often form isolated subnetworks with primarily internal connections.

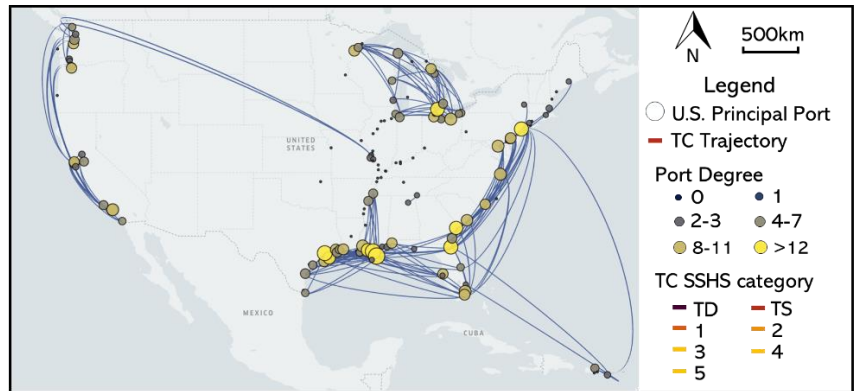

(a) Network During Normal Operations (July 15–21, 2017)

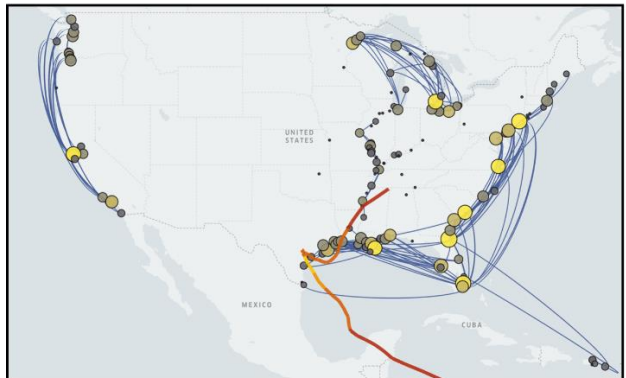

(b) Network Impacted by Hurricane Harvey (August 24–30, 2017)

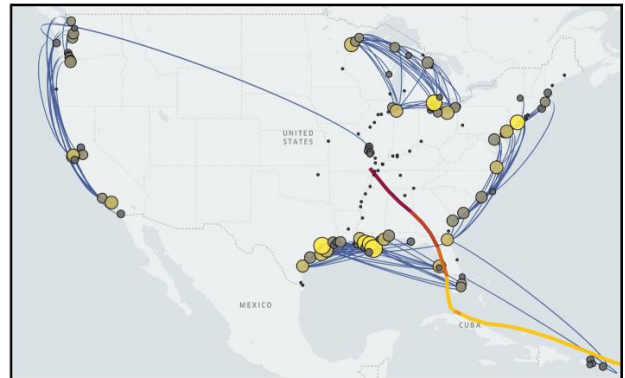

(c) Network Impacted by Hurricane Irma (September 8–14, 2017)

**Fig. 8.** U.S. freight network under normal operations and two common disruption scenarios: (a) a typical week without a tropical cyclone, (b) a week impacted by a tropical cyclone in the Gulf of America, and (c) a week impacted by a tropical cyclone along the East Coast.

The connectivity of the network is highly sensitive to TC disturbances. For example, during Hurricane Harvey's landfall in 2017, which occurred along the Texas Gulf Coast, the weekly freight network displayed a clear fragmentated pattern indicative of disrupted port-to-port linkages. Similarly, Hurricane Irma (2017) offers an illustrative case of East coast disruption. Although Irma's landfall did not directly target Florida's eastern seaboard, major ports such as Port Everglades and Port Canaveral suffered significant declines in connectivity (degree centrality), likely due to strong coastal wind speeds exceeding critical operational thresholds. To further understand the impact, we have modeled the change in node-level connectivity (degree difference) and examined its relationship with centrality-based features. **Table 6** have presented presents results exploring how network position correlates with disruption outcomes.

**Table 6.** Marginal effects for Port's Network Centrality

| Variables | Average Marginal Effects | | |
|---|---|---|---|
| | Total Impact | Recovery Duration | Degree Difference |
| Degree Centrality | - | - | 0.057 |
| Betweenness Centrality | - | - | -8.895 |
| Closeness centrality | 5.637 | 4.451 | 10.603 |

Our results show that port centrality within the freight network plays a critical role in shaping resilience during TC events, with statistically significant coefficients for closeness, betweenness, and degree centrality. As supported by prior studies (Huang et al., 2024), closeness centrality effectively captures the role of transshipment ports, those with high reachability across the network. These ports typically coordinate multi-leg shipping operations, making them essential for maintaining continuity in cargo flows. Our findings indicate a strong positive correlation between closeness centrality and disruption severity across all impact metrics. This suggests that transshipment hubs, due to their network-spanning functions, are more exposed to cascading disruptions when traffic flow is interrupted at multiple points.

In contrast, betweenness centrality, which reflects a node's ability to support alternate shortest paths and serves as a proxy for path redundancy, demonstrates a negative association with degree loss during disruption (-6.570). This finding implies that ports with higher betweenness centrality, acting as flexible routing nodes, tend to experience less severe connectivity degradation. The structural advantage conferred by these ports' integration into multiple shortest paths may allow for rerouting and redistribution of cargo flows when disruptions occur, thereby mitigating overall network disintegration.

Meanwhile, degree centrality, representing the number of direct port-to-port links, is only weakly significant for degree loss. This indicates that while direct connectivity may influence immediate exposure to TCs, it does not independently determine a port's resilience. Instead, it likely interacts with higher-order structural factors such as centrality and network topology.

These findings carry important implications for understanding the resilience of freight networks. While transshipment ports (high closeness) may be more susceptible to amplified disruptions due to their central operational roles, ports with high path redundancy (high betweenness) appear to possess an intrinsic buffering capacity. However, this does not necessarily mean they experience lower absolute impacts or faster recovery — the lack of significance for betweenness in models estimating total impact and recovery duration suggests that their resilience stems more from adaptive rerouting than from insulation against physical impacts.

## 5. Conclusions

In this study, we have presented and analyzed the CyPort dataset, spanning 1,927 port–cyclone interactions across 145 U.S. principal ports and 90 named TCs from 2015 to 2023. Its long time span, integration of multiple data sources, and openly accessible supports future port resilience studies. Building on this dataset, RPNB-Lindley models are applied to examine various effect (e.g. port infrastructure, network connectivity, CT intensities) on port and freight network disruptions. Findings of this study contribute to generalized resilience insights across multiple TCs beyond case-specific studies.

The analysis identifies SSHS Category 4 as a clear resilience tipping point. While lower-category cyclones tend to cause disruptions, category 4 storms trigger sharp increases in vessel activity losses, recovery times, and connectivity breakdowns. This non-linear pattern reflects how impacts escalate disproportionately once storm intensity reaches the threshold of SSHS category 4 . In terms of weather observations at port, , every 1-knot increase in wind speed observation leads to an average of 0.021 additional days of vessel loss and 0.022 more days to recover, while each additional foot of storm surge increases total impact by 0.139 days and causes 0.432 vessel lines to be disconnected. The significance of random parameters associated with tropical cyclone exposure variables further reveals substantial heterogeneity in how ports respond to similar hazard conditions. Therefore, when forecasts indicate an approaching Category 4 or higher storm, port authorities should escalate enhanced emergency measures, such as pre-positioning equipment, securing critical assets, and coordinating with carriers and terminal operators. For storms lower than Category 3, a more targeted response is suggested, depending on the port's specific vulnerabilities and confidence in the forecast trajectory. Early deployment of predictive tools supports proactive planning and reduces the likelihood of cascading disruptions across the freight system.

Meanwhile, port characteristics significantly shape resilience outcomes during tropical cyclones, with coastal ports, especially along the Gulf of America, experiencing the most severe disruptions. Ports on the Pacific Coast have the highest total impact, but it reflects lower impact frequency and differences in preparedness rather than inherent vulnerability. Ports located in more densely populated counties tend to experience less disruption, possibly due to stronger infrastructure and emergency management systems, although their connectivity may remain sensitive due to higher operational complexity. Socioeconomic factors such as poverty rate and workforce composition show weaker associations with impact severity, indicating that meteorological and locational variables are the primary drivers.

In the freight network, ports that serve as intermediary connectors (i.e. high betweenness centrality) better preserves network connectivity under TCs, possibly by supporting alternate vessel routes. In contrast, transshipment hubs with high closeness centrality and local hubs with high degree centrality are more vulnerable, often suffering greater connectivity degradation under TC impact. While these ports are crucial under normal operations, their central roles make them more susceptible to cascading failures during storms. It suggests that transshipment hubs and local hubs should be prioritized in recovery efforts to restore lost capacity and reconnect disrupted flows, while ports with higher betweenness centrality tend to sustain operations on their own by leveraging alternate routing paths within the network.  Based on their vulnerability and role in the network, transshipment and local hubs should be prioritized in post-cyclone recovery to quickly restore capacity and connectivity. By contrast, ports with high betweenness centrality may require less immediate intervention, as they are often able to maintain operations by supporting alternate routing paths.

There are certain limitations in this study. The CyPort dataset captures only vessel movements in U.S. and excludes international routes due to constraints in AIS data coverage, thus potentially underestimating the connectivity of globally oriented ports. Future research could address this by integrating international AIS data (e.g., from Marine Traffic or exact Earth) or incorporating global port connectivity indices.

**Acknowledgement**

This research is funded by the U.S. Department of Transportation through the Center for Freight Transportation for Efficient & Resilient Supply Chain (FERSC). We also gratefully acknowledge Mr. Blakemore of the U.S. Coast Guard (USCG) for sharing his expert insights, which significantly enhanced our understanding of port and freight system resilience.

**Data Availability**

The port–cyclone interaction data are visualized and made accessible through the CyPort Google Earth Engine App (https://CyPort-dataset.users.earthengine.app/view/CyPort).

**Appendix 1. CyPort Dataset Detailed Description**

CyPort emerges as a unified dataset composed of two complementary components: (1) full-period port vessel counts and vessel OD records, and (2) cyclone-specific port-cyclone interaction records. The first component captures fine-grained vessel activity and port operation records, consisting of all port daily vessel count from 2015 to 2023, and 745,704 commercial vessel ODs. Each vessel OD includes the vessel ID, port ID, arrival time, and departure time of both origin and destination port. The second port–cyclone interaction component has 1,927 records, containing descriptors of each interaction's port performance matrices, cyclone profile, weather and tide conditions, and port characteristics. A detailed dataset summary is provided in **Table A-1**.

**Table A-1.** CyPort Data sources Description

| Scope | Data Types | Description | Sources |
|---|---|---|---|
| Baseline Data | Port Daily Vessel Counts | Daily commercial vessel counts for U.S. principal ports from 2015 to 2023 | AIS Data (Marine Cadastre., 2024) |
| | Vessel Trajectories (Port Calls) | Sequential port call records for all commercial vessels (cargo and tanker) | AIS Data (Marine Cadastre., 2024) |
| | Port Characteristics | Includes port statistical areas, trade attributes for 145 principal U.S. ports, along with local census information and connected infrastructure facilities | Port and port statistical area (USAGE, 2021; USDOT BTS, 2022); U.S. Census data (2021); CEJST (2022); Open Street Maps (2024) |
| Cyclone-period Data | Cyclone Records | Tropical cyclone intensity, distance to port, and exposure time window | IBTrACS (NOAA, 2024a) |
| | Meteorological Data | Water level, storm surge height, rainfall and wind speed during TCs, from the nearest gauge and weather stations | CO-OPS (2024); ASOS (NOAA, 2024b) |
| | Port network centralities | Port's degree, betweenness, and closeness centrality in recent weekly freight networks | - |
| | Resilience Metrics | Metrics based on resilience curves and complex network | - |

For cyclone exposure data, we extract TC records from the International Best Track Archive for Climate Stewardship (IBTrACS) (NOAA, 2024a), which includes all tropical cyclone's lasting duration, intensity scales and cyclone eye trajectories. Further, to provide a representation of port-side exposure to cyclone events, storm surge and wind speed were extracted from the nearest CO-OPS gauge stations to ports (CO-OPS, 2024), and total rainfall during the event was obtained from the nearest ASOS weather station (NOAA, 2024b). Following Xu and Huang (2014), maximum surge height was estimated as the difference between peak surge and typical sea level during the TC exposure period, and maximum sustained 6-minute wind speed was used as a measure of wind impact. Port characteristics include port boundary, throughput metrics (e.g., import/export TEUs, port rank), and local resources (e.g., census tract workforce factor, county population) (Council on Environmental Quality, 2022; U.S. Census Bureau., 2021; USDOT BTS, 2022). Additional port connecting facility features such as the number of docks and the total length of highway and railway segments are extracted within a 3 km buffer around each port (OpenStreetMap contributors, 2024; USDOT BTS, 2024). All data features are shown for port-cyclone interactions, as demonstrated in **Table A-2**.

**Table A-2.** Feature reference table for Port-cyclone interaction records

| Category | Feature | Description |
|---|---|---|
| Interaction Identifier | ID | Port-cyclone interaction identifier |
|  | SID | Cyclone Identifier by IBTrACS |
|  | PID | Port identifier |
|  | port | Name of the port |
| Key Time Stamps | start_date | Start date of CT exposure |
|  | end_date | End date of CT exposure |
|  | start_recovery_date | Identified start of recovery date in resilience curve |
|  | end_recovery_date | Identified end of recovery date in resilience curve |
| Resilience Metrics | total_impact | Total impact (unit: equivalent days of recovery) |
|  | total_impact_value | Total impact in vessel count (unit: number of vessels) |
|  | max_impact | Maximum impact |
|  | day_of_recover | Recover duration (unit: days) |
|  | Degree_difference | Degree difference |
|  | CC_difference | Closeness centrality difference |
|  | BC_difference | Betweenness centrality difference |
| Cyclone Exposure | SSHS | the Saffir–Simpson hurricane scale (-1: TD, 0: TS, 1-5: Categories 1-5) |
|  | WIND | CT eye's sustained speed |
|  | PRESSURE | CT eye's pressure |
|  | DISTANCE | CT eyes' closest distance to port (km) |
|  | IF_LANDFALL | Indicator of if CT had landfall in US |
|  | IF_LANDFALL_CLOSE2PORT | Indicator of if CT Landfall location is within 500km to port |
|  | Wind_speed | Sustained wind speed at CO-OPS gauge closest to port (unit: m/s) |
| | Surge_height | Storm surge height at CO-OPS gauge closest to port (unit: m) |
| | Rainfall | Rainfall amount at ASOS station closest to port (unit: mm) |
| Port Characteristics | Port_ranking | Rank of port among 150 principal ports by USAGE |
| | Coast_Gulf_of_Mexico | Indicator of port located in Gulf of Mexico |
| | Coast_Atlantic | Indicator of port located in East Coasst |
| | Coast_Pacific | Indicator of port located in Pacific Coast |
| | Seaport | Indicator of port categorized as seaport |
| | D_normal | Degree of port in recent normal weeks |
| | D_cyclone | Degree of port in the CT impact week |
| | C_normal | Closeness centrality of port in recent normal weeks |
| | C_cyclone | Closeness centrality of port in the CT impact week |
| | B_normal | Closeness centrality of port in recent normal weeks |
| | B_cyclone | Closeness centrality of port in the CT impact week |
| | Pop_C | Port located county population |
| | Pop_Tract | Port located census tract population |
| | PCT_Pov | Percent of individuals below federal poverty line by census tract |
| | WF | Workforce factor by census tract |
| | PCT_TA | Transportation access percentile rank in U.S. by census tract |
| | PCT_ACEW | Anticipated changes in extreme weather percentile rank |
| | PCT_TI | Transportation insecurity percentile rank |
| | PCT_SV | Social vulnerability percentile rank |
| | SL | Sea level rise inundated area indicator |
| | Railway_Length | Port connected railway length within 3km |
| | Highway_Length | Port connected highway length within 3km |
| | Dock_Count | Number of operational docks at the port |

## Appendix 2. Impact Window Detection Algorithm Illustration

Examples of the cyclone impact window detection algorithm are presented in the attached Figure (**Fig. B-1**). The examples cover ports with varying vessel volumes and different cyclone events, showcasing all detected impact windows and key timestamps.

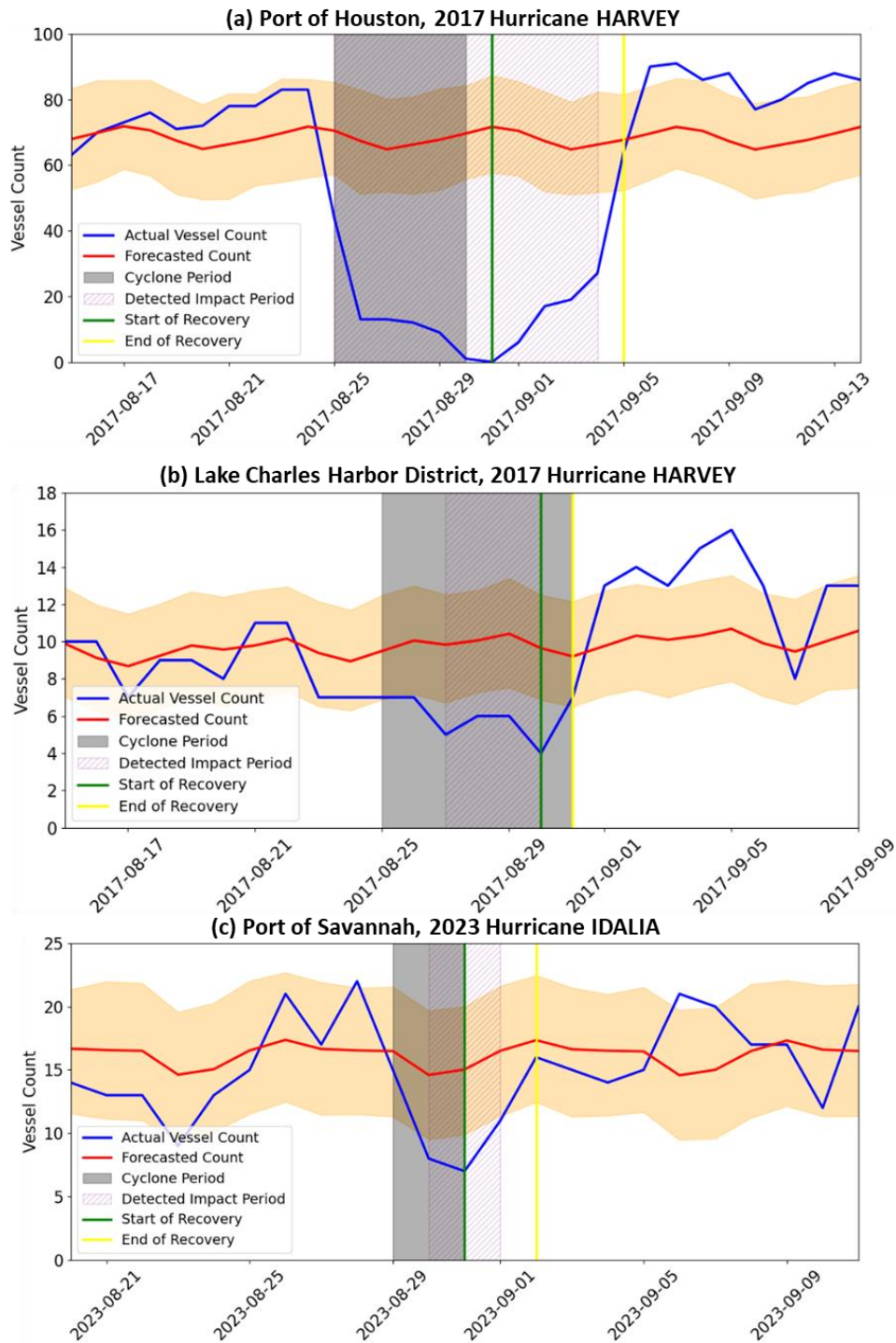

**Fig. B-1.** Performance illustration of the cyclone impact window detection algorithm (across smaller ports and larger ports, different CTs)